\documentclass[english,showpacs,floatfix,12pt]{revtex4}
\usepackage[dvips]{graphicx}
\usepackage{longtable}
\usepackage{amsmath,amssymb}
\usepackage{dcolumn}
\usepackage{latexsym}
\usepackage{babel}
\usepackage[latin1]{inputenc}
\usepackage{color}
\usepackage[colorlinks]{hyperref}
\def\al{\alpha}
\def\kp{\kappa}
\def\nb{\nabla}
\def\pa{\partial}
\def\vf{\varphi}

\def\Om{\Omega}
\def\om{\omega}
\def\Ga{\Gamma}

\def\be{\beta}

\def\Dl{\Delta}

\def\th{\theta}

\def\Sg{\Sigma}

\def\vr{\varrho}
\def\g{{\sqrt g}}
\def\l{\left}
\def\r{\right}
\def\wt{\widetilde}
\def\nn{\nonumber}
\def\diag{\mbox {diag}}
\def\t{\rm {tot}}
\def\U{{\cal {U}}}

\begin{document}

\title{\Large Stationary Spiral Structure and \\
Collective Motion of the Stars in a Spiral Galaxy}
\date{1st May 2009}
\author{Ying-Qiu Gu}
\email{yqgu@fudan.edu.cn} \affiliation{School of Mathematical
Science, Fudan University, Shanghai 200433, China} 
\begin{abstract}
Most fully developed galaxies have a vivid spiral structure, but
the formation and evolution of the spiral structure are still an
enigma in astrophysics. In this paper, according to the standard
Newtonian gravitational theory and some observational facts, we
derive an idealized model for spiral galaxy, and give a possible
explanation to the spiral structure. We solve some analytic
solutions to a spiral galaxy, and obtain manifest relations
between mass density and speed. From the solution we get some
interesting results: (I) The spiral pattern is a stationary or
static structure of density wave, and the barred galaxy globally
rotate around an axis at tiny angular speed. (II) All stars in the
disc of a barred spiral galaxy move in almost circular orbits.
(III) In the spiral arms, the speed of stars takes minimum and the
stellar density takes maximum. (IV) The mass-energy density of the
dark halo is compensatory for that of the disc, namely, it takes
minimum in the spiral arms. This phenomenon might reflect the
complicated stream lines of the dark halo.
\end{abstract}

\maketitle
\section{Introduction}
\setcounter{equation}{0}

Most fully developed galaxies have a vivid spiral structure, which
has profound influences on the generation and evolution of stars.
During the last 60 years many efforts have been made to reveal the
nature of the spiral structure. Despite the progress made in
understanding the amplification mechanisms of the spiral arms and
the role of the spiral arms in the dynamics of galaxies, the
formation and evolution of the spiral structure are still an
enigma in astrophysics\cite{galx0,galx6,sprl}. We even do not know
if the spirals are a long-lived phenomenon, or are short-lived and
regenerated many times during the galactic evolution. C. C. Lin
and F. H. Shu suggested that, the spiral structure should be a
quasi-stationary and stable density waves, which are excited by
over-reflection at co-rotation due to some nonlinear
effects\cite{Lin1,Lin2}. But some other opinions favor short-lived
or recurrent spiral patterns which are developed in the galactic
disks via swing-amplification in a rotating disk with
shear\cite{gold}, or by an external force\cite{Juli}. The recent
observational results seem to favor a long-lived pattern. In
contrast the nearby galaxies with the distant ones observed
through the Hubble Space Telescope and from the
ground\cite{galx2,long1}, the results show that there is only mild
evolution in the relationship between radial size and stellar mass
for galactic disks within the redshift region $0<z<1$.

In the noncentral potential, the orbit of a star is usually an
unclosed and complicated spatial curve, which sensitively depends
on the initial speed and direction\cite{galx0}. So the stellar
system should be more reasonably treated as fluid rather than mass
points. Although the dynamics for the two models is essentially
the same Newtonian mechanics, but the initial and boundary
conditions are different. In the case of fluid, the stream lines
are consistent field, which are a natural result of the generation
of stars from nebulae.

W. Dehnen and J. Binney fitted the mass distribution within the
Milky Way to the observational data such as the rotation curve and
Oort's constants\cite{mass}, but they found the fitting mass
distribution is ill determined. In \cite{galx7,galx8}, the authors
made numerical simulations for 2-dimensional stellar hydrodynamics
of a flat galactic disc embedded in dark matter halo, and the
model solves the Boltzmann equations up to second order
moment\cite{galx0}. A dynamical approach to explain the formation
of both spirals and rings in barred galaxies was proposed in
\cite{ring1,ring2}. It is based on the orbital motion driven by
the unstable equilibrium points of a given rotating bar potential,
and then the spirals, rings and pseudo-rings are related to the
invariant manifolds associated to the periodic orbits around these
equilibrium points. By adjusting dynamical parameters of the host
galaxy, we get the spiral and ring like structures.

If arms of a spiral galaxy were constructed by fixed material, the
arms would become more and more tightly wound, since the matter
nearer to the center of the galaxy rotates faster than the matter
at the edge of the galaxy. The arms would become indistinguishable
from the rest of the galaxy after only a few orbits. In order to
avoid this winding problem, Lin and Shu proposed the density wave
theory\cite{Lin1,Lin2}. They suggested that the arms are not
constructed by stationary materials, but instead made up of areas
of greater density similar to a traffic jam on a highway. More
specifically, for some dynamical reasons, the luminous materials
such as stars intend to move slowly and spend more time inside the
spiral arms and move rapidly outside the spiral arms. Then an
inhomogeneous and quasi-stationary distribution of mass density
wave is developed, which forms the spiral arms. In the sense of
morphology, this theory is quite successful in some aspects.
However, the dynamical explanation for the formation of density
wave is still unclear. The calculations of this paper attempt to
give a clear dynamical explanation for the density wave.

According to the following observational facts, we find the
problem can be simplified and described by an idealized fluid
model, and then the nature of the spiral structure might be
disclosed by dynamical approach.
\begin{enumerate}

\item By observation, except for the center part, all stars in a
spiral galaxy are mainly distributed in a thin disc, so the spiral
structure should be a visible display of collective motion of
these stars, and the collective motion can be approximately
described by the 2-dimensional hydrodynamics.

\item The collision among stars rarely occurs, so the fluid of
star flow is zero-pressure and inviscid, and the stars move along
geodesics.

\item All stars are driven by the average background gravity of
total mass-energy distribution in the galaxy, and the background
mass density and gravity can be constrained by some empirical data
such as the flat rotation curve.
\end{enumerate}
These are some basic assumptions of the following calculations.
The results show that, the stationary solutions to such system
indeed exist, and the distributions of mass density and speed do
have stable spiral structure. So this idealized model and the
solutions might shed some lights on the enigma of galactic
structure.

\section{Hydrodynamics for stars in a spiral galaxy}
\setcounter{equation}{0}

As discussed below, a fully dynamical approach to the galactic
structure in the context of general relativity strongly depends on
the properties of the dark halo, so it is unrealistic at present
due to the lack of such knowledge\cite{galx4,gu1}. Since the
gravity in a galaxy is very weak except for the region near the
center, the Newtonian gravitational theory is accurate enough to
simulate the spiral structure.

Denote the total effective mass-energy density of a galaxy by
$\rho$, the Newtonian gravitational potential by $\Phi$, we have
the dynamical equation for Newtonian gravity (see \cite{wein} or
the appendix)
\begin{eqnarray}
\pa_\al\pa^\al \Phi = - 4\pi G \rho,\label{2.1}
\end{eqnarray}
where $\pa_\al\pa^\al=\pa_t^2-\nb^2$ is the d'Alembert operator.
For the stars moving in the gravity, the dynamical equation of the
flow is given by
\begin{eqnarray}
(\pa_t+\vec V\cdot \nb)\vec V = - \nb \Phi.\label{2.2}
\end{eqnarray}
Theoretically, the complete dynamical equation system should
include the continuity equation and the dynamics of the dark halo.
But these equations strongly depend on the equation of state of
dark matter and dark energy, which is not clearly discovered.
However, as shown bellow, the lack of this knowledge can be
compensated by other empirical data such as the flat rotation
speed curve, and then some important information of galaxy such as
the total mass density and the spiral structure, can be inversely
derived. That is to say, we do the inverse treatments as done in
\cite{mass,ring1,ring2}.

For convenience we rewrite (\ref{2.1}) and (\ref{2.2}) in the
spherical coordinate system $(t,r,\th,\phi)$. By 3-dimensional
tensor operation, we get the following equations
\begin{eqnarray}
c^{-2}\pa_t^2 \Phi -[\pa_r^2+\frac 2 r\pa_r+\frac 1
{r^2}(\pa_\th^2 +\cot\th\pa_\th+\frac 1
{\sin^2\th}\pa_\phi^2)]\Phi+ 4\pi G \rho &=&0,\label{1.1}\\
(\pa_t+V_r\pa_r+V_\th\pa_\th+V_\phi\pa_\phi)V_r-r
V_\th^2-r\sin^2\th V_\phi^2+\pa_r\Phi
&=&0,\label{1.3}\\
(\pa_t+V_r\pa_r+V_\th\pa_\th+V_\phi\pa_\phi)V_\th+\frac
2 r V_r V_\th-\sin\th\cos\th V_\phi^2+\frac {\pa_\th\Phi}{r^2}  &=& 0,\label{1.4}\\
(\pa_t+V_r\pa_r+V_\th\pa_\th+V_\phi\pa_\phi)V_\phi+\frac 2 r V_r
V_\phi + 2\cot\th V_\th V_\phi + \frac {\pa _\phi \Phi}{(r
\sin\th)^2} &=& 0, \label{1.5}
\end{eqnarray}
where the velocities are defined in the form of 3-dimensional
contra-variant vectors
\begin{eqnarray}
V_r =\frac {d r} {dt},\qquad V_\th= \frac {d \th} {dt},\qquad
V_\phi= \frac {d \phi} {dt}. \label{1.6}
\end{eqnarray}

For the stationary spiral structure globally precessing around
$z$-axis at constant angular speed $\Om$ (namely the pattern
speed), under the coordinate transformation $\vf =\phi-\Om t$, the
solution will be static in coordinate system $(t,r,\th,\vf)$,
namely independent of $t$. Then we get the following equations for
such galaxy
\begin{eqnarray}
c^{-2}{\Om^2} \pa_\vf^2 \Phi -[\pa_r^2+\frac 2 r\pa_r+\frac 1
{r^2}(\pa_\th^2 +\cot\th\pa_\th+\frac 1
{\sin^2\th}\pa_\vf^2)]\Phi+ 4\pi G \rho &=&0,\label{1.7}\\
(V_r\pa_r+V_\th\pa_\th+V_\vf\pa_\vf)V_r-r [V_\th^2+\sin^2\th
(V_\vf+\Om)^2]+\pa_r\Phi
&=&0,\label{1.9}\\
(V_r\pa_r+V_\th\pa_\th+V_\vf\pa_\vf)V_\th+\frac
2 r V_r V_\th-\sin\th\cos\th (V_\vf+\Om)^2+\frac {\pa_\th\Phi}{r^2}  &=& 0,\label{1.10}\\
(V_r\pa_r+V_\th\pa_\th+V_\vf\pa_\vf)V_\vf+(\frac 2 r V_r
 + 2\cot\th V_\th )(V_\vf+\Om) + \frac {\pa _\vf
\Phi}{(r \sin\th)^2} &=& 0. \label{1.11}
\end{eqnarray}
In (\ref{1.9})-(\ref{1.11}), some terms have manifest physical
meanings, $r [V_\th^2+\sin^2\th (V_\vf+\Om)^2]$ stands for
centrifugal force, and $(\frac 2 r V_r
 + 2\cot\th V_\th )\Om$ the Coriolis force.

Considering an unwarped galaxy with two stationary spiral aims,
the total mass-energy density and potential can be generally
expanded by spherical harmonics $Y_{lm}(\th,\vf)$ with even $m$.
To second order terms, we equivalently have the following
approximation
\begin{eqnarray}
\rho &=& \rho_0+(\rho_1 + \rho_2\cos2\vf +\rho_3  \sin2\vf) \sin^2\th,\label{1.12}\\
\Phi &=& \Phi_0+(\Phi_1+ \Phi_2\cos2\vf +\Phi_3  \sin2\vf)
\sin^2\th,\label{1.13}
\end{eqnarray}
where all $(\rho_n,\Phi_n)$ are functions of $r$, and
$(\rho_0\ge0,\rho_1 \ge 0)$. Accordingly, the velocity of the
stars in the disc, also to second order terms,  should be
\begin{eqnarray}
V_r &=& W_1 \cos2\vf +W_2 \sin2\vf , \qquad V_\th=0, \label{1.14}\\
V_\vf &=& \om_0+\om_1 \cos2\vf +\om_2 \sin2\vf ,\label{1.15}
\end{eqnarray}
where all $(W_n,\om_n)$ are functions of $r$.

Substituting (\ref{1.12}) and (\ref{1.13}) into (\ref{1.7}), we
get the relations between $\Phi_n$ and $\rho_n$ as follows
\begin{eqnarray}
\pa_r^2 \Phi_0+\frac 2 r\pa_r \Phi_0+\frac 4 {r^2}\Phi_1 &=& 4\pi G\rho_0,\label{1.16}\\
\pa_r^2 \Phi_1+\frac 2 r\pa_r \Phi_1-\frac 6 {r^2}\Phi_1 &=& 4\pi
G\rho_1,\label{1.17} \\
4c^{-2}\Om^2 \Phi_k+\pa_r^2 \Phi_k+\frac 2 r\pa_r \Phi_k-\frac 6
{r^2}\Phi_k &=& 4\pi G\rho_k,~~(k=2,3). \label{1.18}
\end{eqnarray}
Substituting (\ref{1.13})-(\ref{1.15}) into (\ref{1.9}) and
(\ref{1.11}), and constraining $\th=\frac 1 2\pi$, we get
respectively
\begin{eqnarray}
0 &=& \frac 1 4 \pa_r (W_1^2+W_2^2)+\pa_r(\Phi_0+\Phi_1)-\frac 1 2
r
(2\om_0^2+\om_1^2+\om_2^2)+\om_1 W_2-\om_2 W_1+\nn\\
&~&2[(\om_0 -\Om) W_2-r\om_0\om_1+\frac 1 2 \pa_r\Phi_2]\cos2\vf
+\nn\\
&~&2[(\Om -\om_0) W_1-r\om_0\om_2+\frac 1 2 \pa_r \Phi_3]\sin2
\vf+\cdots, \label{1.19}
\end{eqnarray}
and
\begin{eqnarray}
0&=&\frac1 2 (r \pa_r \om_1+2\om_1)W_1+\frac 1 2
(r\pa_r\om_2+2\om_2)W_2+\nn\\
&~&[(r\pa_r\om_0+2\om_0)W_1+2(\om_0-\Om)r\om_2+\frac 2 r
\Phi_3]\cos2\vf+\nn\\
&~&[(r\pa_r\om_0+2\om_0)W_2+2(\Om-\om_0)r\om_1-\frac 2 r
\Phi_2]\sin2\vf+\cdots.\label{1.20}
\end{eqnarray}
The equation (\ref{1.10}) automatically holds. Theoretically, we
can solve $(\Om, W_j, \om_j)$  and some relations among
$(\rho_k,\Phi_k)$ by condition that the coefficients of
(\ref{1.19}) and (\ref{1.20}) vanish.

According to the high-accurate observational
data\cite{galx1,rotat1,rotat2}, we find the rotation curves of
most spiral galaxies are approximately flat, then we can assume
\begin{eqnarray}
\om_0=\frac v r - \Om, \qquad (r\in [R_0, R_1 ]), \label{3.1}
\end{eqnarray}
where $v$ is a constant speed with typical value $|v|=200\sim
400$km/s, $[R_0,R_1 ]$ is the effective region of (\ref{3.1}),
with their typical values as $R_0= 100\sim500$pc, and $R_1
=10\sim60$kpc to be approximately the visible radius of the
galaxy. Equivalently, we can assume $(\om_0>0, v>0, \Om>0)$ in
calculation, because the dynamical equations
(\ref{1.1})-(\ref{1.5}) have reversal invariance under
transformation $\phi\to-\phi$. We use the empirical condition
(\ref{3.1}) to replace the dynamical equations for the background,
then the problem is greatly simplified. In addition, for a given
galaxy, instead of constant $v$, we can use fitting function
$v=v(r)$ in (\ref{3.1}) to get more accurate results and larger
effective domain\cite{fit1,fit2,fit3}. However in this case, we
can usually get numerical results only.

Substituting (\ref{3.1}) into (\ref{1.19}) and (\ref{1.20}), by
setting the coefficients of $(\sin2\vf,\cos2\vf)$ terms to zero,
we can solve
\begin{eqnarray}
\om_1 &=& -K[\frac 1 2v\pa_r \Phi_2+\frac 2 r (v-r\Om)\Phi_2], \label{3.2} \\
\om_2 &=& -K[\frac 1 2v\pa_r \Phi_3+\frac 2 r (v-r\Om)\Phi_3], \label{3.3} \\
W_1 &=& K[(v-r\Om)r\pa_r\Phi_3+2v\Phi_3], \label{3.4} \\
W_2 &=& K[(v-r\Om)r\pa_r\Phi_2+2v\Phi_2], \label{3.5}\\
K &\equiv & (2r^2\Om^2-4rv\Om+v^2)^{-1}. \label{3.6}
\end{eqnarray}
By the zeroth order terms in (\ref{1.19}) and (\ref{1.20}), we
have
\begin{eqnarray}
\pa_r(\Phi_0+\Phi_1)+\frac 1 4 \pa_r (W_1^2+W_2^2)=\frac 1 2 r
(2\om_0^2+\om_1^2+\om_2^2)-\om_1 W_2+\om_2 W_1, \label{2.27}
\end{eqnarray}
and
\begin{eqnarray}
(r \pa_r \om_1+2\om_1)W_1+(r\pa_r\om_2+2\om_2)W_2=0. \label{2.28}
\end{eqnarray}
Substituting (\ref{3.2})-(\ref{3.6}) into (\ref{2.27}) and
(\ref{2.28}), we get two constraints for $\Phi_k(r)$ of the
background potentials.

Obviously, the disc satisfies the mass conservation law
independent of the dark halo, so we have the 2-dimensional
continuity equation for all stars and baryonic particles moving in
the disc as follows\cite{sprl,galx7}
\begin{eqnarray}
0 &=& \pa_t\Sg  +\nb\cdot(\vec V \Sg ), \nn\\
&=& (\pa_t+V_r\pa_r+V_\phi\pa_\phi)\Sg +(\pa_r V_r+\pa_\phi
V_\phi+\frac 1 r V_r)\Sg , \label{1.21}
\end{eqnarray}
where $\Sg $ stands for the surface mass density of the stars and
baryons in the disc. It should be mentioned that,
(\ref{1.7})-(\ref{1.11}) are expressed in spherical coordinate
system due to the background gravity, but (\ref{1.21}) is
expressed in polar coordinate system. In the stationary case,
(\ref{1.21}) becomes
\begin{eqnarray}
(V_r\pa_r+V_\vf\pa_\vf)\Sg +(\pa_r V_r+\pa_\vf V_\vf+\frac 1 r
V_r)\Sg =\nb\cdot(\vec V \Sg )=0. \label{1.22}
\end{eqnarray}
(\ref{1.21}) or (\ref{1.22}) is the equation to describe the mass
density of the stars.

\section{resolution to the equations}
\setcounter{equation}{0}
\subsection{Solution to a Barred Spiral Galaxy}
The general solutions to the above underdetermined equation system
are quite complicated and unnecessary. The most important case to
understand the nature of spiral structure is the stable and
terminal state of a galaxy, which is similar to the eigenstate of
a micro particle. At first, we consider the case that all stars
move in the orbits near circle, for which the analytic solutions
can be solved. In this case the radial speed is a high order
little term, then we have
\begin{eqnarray}
W_1=0,\qquad W_2=0. \label{3.7}\end{eqnarray} By (\ref{3.4}) and
(\ref{3.5}), noticing the symmetry between $(\Phi_2,\Phi_3)$, we
get the equivalent solution, except for an initial phase of $\vf$,
as follows
\begin{eqnarray}
\Phi_2 =- \frac {q}{r^2}(v-r\Om)^2,\qquad \Phi_3=0,
\label{3.8}\end{eqnarray} where $q\ge0$ is a constant.
Substituting (\ref{3.7}) and (\ref{3.8}) into the above equations
(\ref{1.16})-(\ref{1.18}) and (\ref{3.2})-(\ref{2.28}), we finally
get
\begin{eqnarray}
\om_1 &=& \frac {q} {r^3}(v-r\Om),\qquad \om_2=0,\qquad \rho_3=0,\label{3.9} \\
\rho_1 &=&-\rho_0+ \frac 1{4\pi G}\l( \frac
{q^2v}{12r^6}(16r\Om-15v)-\frac {2v^2} {r^2}\ln\frac r{r_0}+\frac
{2\Phi_0}{r^2}+\frac {v^2}{r^2} \r),\label{3.04}\\
\rho_2 &=&\frac {{q}}{2\pi G{r^4}}\l ((3r^2\Om^2-6rv\Om
+2v^2)-\frac
{2r^2\Om^2}{c^2}(v-r\Om)^2\r),\label{3.11} \\
0 &=&\Phi_0 +\Phi_1 + \frac {q^2}{24r^4}(6r^2\Om^2-8vr\Om+3v^2) -
v^2\ln\frac r {r_0},\label{3.06}
\end{eqnarray}
where $r_0>0$ is a constant with length dimension.

Substituting the solutions into (\ref{1.9}) and (\ref{1.11}), we
can check the truncation error of the equations reads
\begin{eqnarray}
\Dl (\ref{1.9}) =- \frac {q^2} {2r^5} (v-r\Om)^2 \cos 4\vf,\quad
\Dl (\ref{1.11}) = - \frac {q^2} {r^6} (v-r\Om)^2\sin
4\vf,\label{3.13}
\end{eqnarray}
which are higher order terms. So the above solution is a good
approximation. If $q=0$, the solutions are exact, which correspond
to the galaxy without spirals in the disc. To get more accurate
results, one should introduce higher order terms such as $(\cos
4\vf,\sin 4\vf)$ terms in (\ref{1.12})-(\ref{1.15}).

By (\ref{3.1})-(\ref{3.8}), we get the speed of the stellar flow
\begin{eqnarray}
V_r = 0, \qquad V_\th =0,\qquad V_\vf =\frac v r+\frac {q}{r^3}
(v-r\Om)\cos2\vf.\label{3.14}
\end{eqnarray} Substituting it into
(\ref{1.22}), we get the continuity equation as $\pa_\vf(V_\vf \Sg
)=0$, which yields the mass density distribution of the stars as
\begin{eqnarray}
\Sg (r,\vf)= \frac
{vr^2\vr(r)}{vr^2+q(v-r\Om)\cos2\vf},\label{rho}
\end{eqnarray}
where $\vr(r)$ is a density function determined by boundary
conditions. Apparently, the density distribution of the stars
display a profile of barred spiral galaxy.

To determine $(\rho_0,\rho_1)$, we need another condition related
to the dynamics of the black halo. In the case of $\rho_1\propto
\rho_0$, we can solve
\begin{eqnarray}
\rho_0 &\sim& \frac {n^2-n-4}{4\pi G} q^2\l(\frac
{9v^2}{2(n-4)(n+3)r^6}-\frac{\Om^2}{(n-2)(n+1)r^4}\r)+\nn\\
&~&\frac {n^2-n-4}{4\pi G} \l(\frac a {r^{2+n}}+\frac
{3v^2}{n(n-1)r^2}\r) ,\label{3.15}\\
\rho_1 &\sim& \frac{(3-n)(n-2)}{n^2-n-4}\rho_0,\label{3.16}
\end{eqnarray}
in which $a$ and $\frac 1 2(\sqrt{17}+1)\le n\le 3$ are constants.
(\ref{3.15}) and (\ref{3.16}) provide a heuristic mass density for
the dark halo in barred spiral galaxy within the range
$[R_0,R_1]$.

\subsection{Solution to a Spiral Galaxy}
In an ordinary spiral galaxy, the potentials $\Phi_2$ and $\Phi_3$
should take the following form
\begin{eqnarray}
\Phi_2 =P^{-1}\cos (\xi r+\vf_0),\qquad \Phi_3=P^{-1}\sin (\xi
r+\vf_0), \label{B.1}\end{eqnarray} where $\xi$ is a constant,
$\vf_0$ the initial phase of $\vf$ and $P(r)$ a function of $r$.
Obviously we can set $\vf_0=0$ by a translation of $\vf$ similar
to the barred spiral case. Substituting (\ref{B.1}) into
(\ref{2.28}), we get a linear equation for $P$
\begin{eqnarray}
P''-\frac {4v}{r(v-r\Om)} P' +
\l(\xi^2+\frac{4\Om^2}{v(v-r\Om)}+\frac {4\Om}{rv}+\frac
8{r^2}\r)P=0. \label{B.2}\end{eqnarray} The equation (\ref{B.2})
is independent of $\cos (\xi r+\vf_0)$ and $\sin (\xi r+\vf_0)$,
which reflects (\ref{B.1}) touches the nature of the spiral
structure.

The solution to (\ref{B.2}) can be expressed by the Hankel
function with complex parameters, which is much complicated. Here
we examine the simplest case of (\ref{B.2}), the static case with
$\Om=0$. The analytic solution to this case can be expressed in a
clear form, and it is heuristic to reveal the nature of a spiral
galaxy. In the general cases, the numerical simulation is more
convenient and efficient.

Setting $\Om=0$, (\ref{B.2}) becomes a Bessel-like equation
\begin{eqnarray}
P''-\frac {4}{r} P' + (\xi^2+\frac 8{r^2})P=0.
\label{B.2.1}\end{eqnarray} The solution to (\ref{B.2.1}) is given
by
\begin{eqnarray}
P=\sqrt{r^5}\l[C_1 J_\al(|\xi| r)+C_2 J_{-\al}(|\xi| r)\r],\qquad
\al= \frac 1 2 {\sqrt 7}, \label{B.3}\end{eqnarray} where $(C_1,
C_2)$ are constants, $(J_\al,J_{-\al})$ are Bessel functions with
complex parameters, which are defined by
\begin{eqnarray}
J_\nu(x)=\l(\frac x 2\r)^\nu\sum_{k=0}^\infty\frac 1 {k!}\frac 1
{\Ga(\nu+k+1)}\l(\frac x 2\r)^{2k},~~~(\nu=\pm \al).
\label{B.4}\end{eqnarray} Substituting (\ref{B.3}) into
(\ref{1.18}) and (\ref{3.2})-(\ref{3.6}), we can check
$(\rho_2,\rho_3)$ and speed all have spiral structure like
(\ref{B.1}). Again by (\ref{1.22}), we find $\Sg$ also has spiral
structure. Substituting the results into (\ref{2.27}), we get a
constraint for $(\Phi_0,\Phi_1)$ similar to (\ref{3.06}). Since
the concrete expressions are long and complicated, and can be
derived by straightforward calculation, we do not display them
here.

\section{discussion and conclusion}
\setcounter{equation}{0}

In the context of general relativity, the whole dynamical equation
system for the galactic evolution should be the Einstein's field
equation
\begin{eqnarray}
G^{\mu\nu}\equiv R^{\mu\nu}-\frac 1 2 g^{\mu\nu} R=-\kp T^{\mu
\nu},\qquad (\kp\equiv \frac{8\pi G} {c^4}), \label{g.1}
\end{eqnarray}
combined with the energy-momentum conservation law and the
equation of state of the gravitating source. For the spinors with
interactions, the classical approximation gives\cite{gu1}
\begin{eqnarray}
T^{\mu\nu}=(\rho_{\t} +P)\U^\mu \U^\nu +(
W-P)g^{\mu\nu},\label{g.2}
\end{eqnarray}
where $W$ is a potential corresponding to the interaction terms,
which acts as negative pressure. According to the energy-momentum
conservation law or Bianchi identity $T^{\mu\nu}_{~;\nu}=0$, we
can derive the continuity equation $\U_{\mu} T^{\mu\nu}_{~;\nu}=0$
and the equation of motion for the source as follows
\begin{eqnarray}
\U^\mu \pa_\mu (\rho_{\t} +W)&=&-( \rho_{\t} +P)\U^\mu_{~;\mu}, \label{g.3}\\
(\rho_{\t} +P) \U^{\nu} \U^\mu_{~;\nu}&=&(g^{\mu\nu}-\U^\mu
\U^{\nu})\pa_\nu (P-W). \label{g.4}
\end{eqnarray}
For the nonlinear dark spinors, we have $W\sim \rho_{\t}\gg P$. In
this case, by (\ref{g.4}) we find the stream lines of the spinors
are quite different from the geodesics $\U^\nu \U^\mu_{~;\nu}=0$.
So unless the nature of the dark matter is disclosed, a fully
relativistic simulation for the dynamics of galaxy is impossible.
In (\ref{2.1}), the effects of $(P,W)$ are merged into one
effective mass-energy density $\rho$, so the treatment is much
simplified.

The above procedure provides a method to research the structure of
a galaxy via dynamical approach, which naturally connect the
hydrodynamics with empirical data. The above solutions provide
manifest analytic relations and functions for a spiral galaxy,
which are much helpful to understand the structure and property of
the galaxy. These solution verifies that the spiral arms are
stationary or even static stellar density wave distribution.

In the case of barred spiral galaxy with $\Om\ne0$, by
(\ref{3.11}), we find $\rho_2(r)= 0$ has a root $r=R \lesssim
v/\Om$, which means the bar vanishes at radius $r=R \sim R_1 $. By
(\ref{3.9}), we get $\om_1(R)\dot= 0$, which implies the density
wave vanishes near $r=R$, and then a ring of stars will form. The
above solution is derived under the assumption (\ref{3.1}), so the
effective domain of the solution is also $[R_0,R_1 ]$, the
effective domain of (\ref{3.1}). These conclusions can be used to
estimate the global angular speed
$$ \Om\approx \frac v R_1  \sim 300{\rm km/s}
/(30{\rm kpc})= 10{\rm km/s/kpc} \sim 0.002''/{\rm year},$$ which
is less than the previously estimated pattern speed $30\sim 60
{\rm km/s/kpc}$\cite{galx9}.

By (\ref{3.13}) and (\ref{3.14}), we learn all stars in the disc
move in the almost circular orbits. This conclusion is reasonable
and coincident with facts. Through some mechanisms and long-term
evolution, the baryons were shifted and concentrated into the
disc, and their orbits of motion became harmonious in a regular
galaxy. Otherwise violent collisions will occur frequently among
the moving stars, and then the galaxy become a hell.

By (\ref{3.9}) and (\ref{3.11}), we find $\om_1>0$ and $\rho_2>0$
hold simultaneously. This means where the mass density of the
background is larger, where the potential is lower and the stellar
speed is higher. On the other hand, by (\ref{rho}) we find that
the higher the stellar speed, the lower the number density. This
means the mass density of the disc should be compensatory for that
of the dark halo. Namely, the density of dark halo $\rho$ should
take minimum in the spiral arms where $\Sg $ take maximum or vice
versa. This is a strange but interesting phenomenon, which
reflects the complicated stream lines of the dark halo. This
phenomenon might also be one reason why the dark matter and dark
energy can be hardly detected in the vicinity of the solar system.

The relations (\ref{3.7}) and (\ref{3.04})-(\ref{3.06}) as well as
(\ref{B.1}) provide some information for the distribution and
property of the total mass density and potential. How to combine
the results with the dynamics of the background is an interesting
problem, which might be a shortcut to study the properties of the
dark matter and dark energy.
\section*{Acknowledgments}
The appendix is added according to one referee's suggestion of
MNRAS. The author is grateful to his supervisors Prof. Ta-Tsien Li
for his encouragement.

\section{Appendix: The Derivation of the Equations (\ref{2.1}) and (\ref{2.2})}
\setcounter{equation}{0}

In this appendix, we derive the dynamics (\ref{2.1}) and
(\ref{2.2}) of the Newtonian gravitational system from the
Einstein's relativistic dynamics (\ref{g.1})-(\ref{g.4}) by
weak-field and low-speed approximation. Some fundamental contents
can be found in \cite{wein}, but here we make more systematic and
detailed study for galactic dynamics. For convenience, we take
$c=1$ as unit of velocity. Noticing the facts that the collisions
among stars rarely occur, and the trajectories of the ordinary
matter such as electrons and baryons are almost geodesics, so for
the stars, the following zero-pressure and inviscid
energy-momentum tensor holds
\begin{eqnarray}
T^{\mu\nu}_{s}=\rho_s U^\mu U^\nu,\label{a1.2}
\end{eqnarray}
in which $\rho_s$ is the comoving mass density of the stars, and
$U^\mu$ is the 4-vector speed of the stellar flow. Since the
ordinary matter satisfies the mass-energy conservation law
independent of the dark halo, we have $T^{\mu\nu}_{s;\nu}=0$.
Expressing it in the form of equations of continuity and motion,
we get the dynamical equations for the stars
\begin{eqnarray}
U^\mu \pa_\mu \rho_s + \rho_sU^\mu_{~;\mu}=0,\qquad U^{\nu}
U^\mu_{~;\nu}=0. \label{a1.3}
\end{eqnarray}

The total energy-momentum tensor of the galaxy is still given by
(\ref{g.2}), and satisfies the dynamical equations (\ref{g.3}) and
(\ref{g.4}). By (\ref{g.1}) and (\ref{g.2}), we get
\begin{eqnarray}
R= \kp(\rho_{\t}+4W-3P),\label{a1.14}
\end{eqnarray}
where $R=g_{\mu\nu}R^{\mu\nu}$ is the scalar curvature.
Substituting (\ref{a1.14}) into (\ref{g.1}), we get
\begin{eqnarray}
R^{\mu\nu}=-\kp(\rho_{\t}+P) \U^\mu \U^\nu+\frac 1 2
\kp(\rho_{\t}+2W-P) g^{\mu\nu}, \label{a1.15}
\end{eqnarray}
where $\U^\mu$ is the average 4-vector speed of all gravitating
source.

In order to make weak-field approximation, we choose the harmonic
coordinate system, which leads to usual Cartesian coordinate
system when making linearization of metric. Then we have the de
Donder coordinate condition
\begin{eqnarray}
\Ga^\mu \equiv g^{\al\be}\Ga^\mu_{\al\be}=-\frac 1 \g \pa_\nu(\g
g^{\mu\nu})=0, \label{a1.16}
\end{eqnarray}
where $g={|\det(g)|}$. Denote the Minkowski metric by
\begin{eqnarray}
\eta_{\mu\nu}=\eta^{\mu\nu}=\diag(1, -1, -1, -1). \label{a2.1}
\end{eqnarray}
For weak-field approximation, we have the linearization for the
metric
\begin{eqnarray}
g_{\mu\nu}&\equiv &\eta_{\mu\nu}+h_{\mu\nu},\quad
g^{\mu\nu}\dot =\eta^{\mu\nu}-h^{\mu\nu},\label{a2.3}\\
h^{\mu\nu}&=&\eta^{\mu\al}\eta^{\nu\be}h_{\al\be},\quad~
h=h^\mu_{~\mu}=\eta^{\mu\nu}h_{\mu\nu},\label{a2.2}\\
g &\dot =& 1+h,\qquad\quad  \g \dot = 1+\frac 1 2 h.\label{a2.4}
\end{eqnarray}
In what follows, we directly use $=$ to replace $\dot =$. By
straightforward calculation, we get the linearization for other
parameters
\begin{eqnarray}
\Ga^\mu_{\al\be} &=& \frac 1 2\eta^{\mu\nu}(\pa_\al h_{\nu\be}+\pa_\be h_{\al\nu}-\pa_\nu h_{\al\be}),\label{a2.5}\\
\Ga^\mu&=& \pa_\nu (h^{\mu\nu}-\frac 1 2 \eta^{\mu\nu}h), \label{a2.6} \\
R_{\mu\nu}&=&\frac 1 2 \pa_\al\pa^\al h_{\mu\nu}-\frac 1 2
(\eta_{\mu\al}\pa _\nu \Ga^\al
+ \eta_{\nu\al}\pa _\mu \Ga^\al) \label{a2.7},\\
R^{\mu\nu}&=&\frac 1 2 \pa_\al\pa^\al h^{\mu\nu}-\frac 1 2
(\eta^{\mu\al}\pa _\al \Ga^\nu+\eta^{\nu\al}\pa _\al \Ga^\mu),\label{a2.8}\\
R&=&\frac 1 2 \pa_\al\pa^\al h -\pa _\al \Ga^\al. \label{a2.9}
\end{eqnarray}
In the harmonic coordinate system, we have
\begin{eqnarray}
\Ga^\mu &=& \pa_\nu ( h^{\mu\nu}-\frac 1 2 \eta^{\mu\nu} h)=0, \label{a2.10} \\
R_{\mu\nu}&=&\frac 1 2 \pa_\al\pa^\al h_{\mu\nu},~~~
R^{\mu\nu}=\frac 1 2 \pa_\al\pa^\al h^{\mu\nu}, \label{a2.11}\\
R&=&\frac 1 2 \pa_\al\pa^\al h,~~~G^{\mu\nu} = \frac 1 2 \pa_\al
\pa^\al(h^{\mu\nu}-\frac 1 2 \eta^{\mu\nu} h).\label{a2.12}
\end{eqnarray}
By (\ref{a2.10}) and (\ref{a2.12}) we find if $\Ga^\mu=0$ at any
given time $t=t_0$, it will always hold due to the Bianchi
identity $G^{\mu\nu}_{~;\nu}=0$.

In order to compare with electromagnetism and to understand the
physical meaning of the parameters, denote
\begin{eqnarray}
{\Phi} &=&\frac 1 2 h_{tt}=\frac 1 2 h^{tt},~~\vec A = (h^{tx},h^{ty},h^{tz})=-(h_{tx},h_{ty},h_{tz}), \label{a2.13}\\
H &=& (h_{ab})= (h^{ab}),~(\{a,b\} \in \{1,2,3\}),\quad \vec B =
\nb \times \vec A.\label{a2.14}
\end{eqnarray}
In the International System of Units, we have the order of
magnitude for the components of metric
\begin{eqnarray}
c^2|h_{ab}|\sim  c |A_k|\sim |\Phi|\ll 1 ,~~(a\ne b),
\end{eqnarray}
which means $|h_{ab}|\ll|A_k|\ll|\Phi|\ll 1$ if taking $c=1$ as
unit.

For the present purpose, we define the stellar speed $\vec V$ by
\begin{eqnarray}
\vec V\equiv \frac 1{U^0}(U^1,U^2,U^3), \label{a2.16}
\end{eqnarray}
which is approximately equivalent to the usual definition. For
galaxies, we have the following order of magnitude
\begin{eqnarray}
|\vec V|\sim 300{\rm km/s} =10^{-3} c,\quad \vec A\sim \kp \vec
V,\quad h_{ab}\sim \kp |\vec V|^2,~(a\ne
b),\label{5.22}\end{eqnarray} in which the coefficient $\kp$ is
also a number of little value. Then according to
\begin{eqnarray} 1=\sqrt{g_{\mu\nu} U^\mu U^\nu}= ( 1+2 \Phi -2\vec A
\cdot \vec V+g_{ab} V^a V^b )^{\frac 1 2} U^0,\label{a2.17}
\end{eqnarray}
by omitting $O(V^2)$ terms, the low-speed assumption gives
\begin{eqnarray} U^0 = 1 - \Phi +\vec A
\cdot \vec V.\label{a2.18}
\end{eqnarray}
Substituting (\ref{a2.16}) and (\ref{a2.18}) into (\ref{a1.3}) and
omitting the high order terms, we get the continuity equation and
motion equation for stars
\begin{eqnarray}
(\pa_t+\vec V\cdot \nb)\rho_s &=& - \rho_s [ \nb\cdot \vec V+
(\pa_t\Phi+\nb\cdot \vec A)], \label{a2.19}\\
(\pa_t+\vec V\cdot \nb)\vec V &=& - \nb \Phi+(-\pa_t \vec A + \vec
V\pa_t \Phi)+ \vec V\times \vec B+\vec V\cdot\pa_t H.\label{a2.20}
\end{eqnarray}
In (\ref{a2.19}), we used the de Donder condition $\Ga^0=0$ in the
form
\begin{eqnarray}
\frac 1 2 \pa_t (h_{xx}+h_{yy}+h_{zz})=-(\pa_t\Phi+\nb\cdot \vec
A).
\end{eqnarray}

The equation of motion (\ref{a2.20}) has a similar structure to
the electrodynamics. From it we learn that, $\Phi$ gives the
Newtonian gravitational potential, and $\vec A$ leads to
gravimagnetic field $\vec B$. By (\ref{5.22}), the zeroth order
approximations of (\ref{a2.19}) and (\ref{a2.20}) are just
(\ref{1.21}) and (\ref{2.2}) respectively.

By (\ref{a1.14}) and (\ref{a2.12}), we have
\begin{eqnarray}
\pa_\al\pa^\al h=2\kp(\rho_{\t}+4W-3P). \label{a2.22}
\end{eqnarray}
By (\ref{a2.22}), (\ref{a1.15}) and (\ref{a2.11}), we get the
dynamical equations for $h^{\mu\nu}$
\begin{eqnarray}
\pa_\al\pa^\al h^{\mu\nu} &=& -2\kp(\rho_{\t}+P) \U^\mu
\U^\nu + \kp (\rho_{\t} + 2W-P) \eta^{\mu\nu}, \label{a2.24}\\
\pa_\al\pa^\al \chi^{\mu\nu} &=& -2\kp[(\rho_{\t}+P) \U^\mu \U^\nu
+(W-P) \eta^{\mu\nu}], \label{a2.23}
\end{eqnarray}
where $\chi^{\mu\nu} =h^{\mu\nu}-\frac 1 2 \eta^{\mu\nu}h$. If the
average speed of the dark halo is also small, omitting $O(\vec
\U^2)$ from (\ref{a2.24}) we get $h^{xx}=h^{yy}=h^{zz}\equiv 2
\Psi$, $h^{ab}=0,(a\ne b)$ and
\begin{eqnarray}
\pa_\al\pa^\al \Phi &=& \frac 1 2 \pa_\al\pa^\al h^{00}=-4\pi
G\rho,\label{a2.26} \\
\pa_\al\pa^\al \Psi &=& \frac 1 2 \pa_\al\pa^\al h^{kk}=-4\pi G\wt
\rho,\label{a2.27}
\end{eqnarray}
where $\rho$ and $\wt\rho$ are the effective mass density. Their
zeroth order approximation gives
\begin{eqnarray}
\rho&=&\rho_{\t}[2(\U^0)^2-1]-2W+P[2(\U^0)^2+1]\dot =
\rho_{\t}-2W+3P.\label{5.32}\\
\wt\rho&\dot =& \rho_{\t}+2W-P.\label{5.33}
\end{eqnarray}
The equation (\ref{a2.26}) is just (\ref{2.1}). For the dark
matter or dark energy with large enough negative pressure, by
(\ref{5.32}) we may even have $\rho<0$, which means the Newtonian
gravity becomes repulsive in this case. So detecting the dynamical
behavior of the galactic dark halo may be a shortcut to
investigate the weird properties of dark matter or energy.

\end{document}